\documentclass{epl}

\title{A Damping of the de Haas-van Alphen Oscillations in the superconducting state}
\author{K. P. Duncan \and B. L. Gy\"orffy}
\institute{
  H. H. Wills Physics Laboratory, University of Bristol, Tyndall Avenue, Bristol BS8 1TL, UK.
}
\begin{document}

\maketitle

\begin{abstract}
Deploying a recently developed semiclassical theory of quasiparticles in the 
superconducting state we study the de Haas-van Alphen effect. We find that the
 oscillations have the same frequency as in the normal state but their 
amplitude is reduced. We find an analytic formulae for this damping which is
due to tunnelling between semiclassical quasiparticle orbits comprising both 
particle-like and hole-like segments. The quantitative predictions of the theory are consistent with the available data.
\end{abstract}

The revival of interest \cite{onuki:92:0,janssen:98:0,goll:96:0} 
in studying the de Haas-van Alphen 
(dHvA) effect in the superconducting state \cite{graebner:76:0} is driven by the hope that
this would provide new ${\bf k}$-vector dependent information about the 
superconducting gap $\Delta({\bf k})$. Evidently this would be of particular
importance in connection with anisotropic superconductors where 
$\Delta({\bf k})$ can have lines of zero's on the Fermi surface 
\cite{mineev:99:0}.
Unfortunately at this stage there is no concensus concerning the mechanism of
how the experimentally observed oscillations of the diamagnetic response of 
a Type II superconductor come about 
\cite{gorkov:98:0,maniv:01:0,miller:95:0,miyake:93:0,dukan:95:0,norman:95:0,wasserman:96:0,burmistrov:96:0,bruun:97:0,vavilov:97:0}. 
Using our recently developed very general semiclassical theory of 
quasiparticles in the superconducting state \cite{duncan:02:0}, 
in what follows we develop a semiclassical picture of
Landau like orbits of quasiparticles suggested by the simple model calculation
of Miller and Gy\"orffy \cite{miller:95:0}. 
Clearly the long term aim of a semiclassical
theory is to provide an analogue of the Lifshitz-Kosevich formulae for 
superconductors. Hopefully such a formulae would allow the interpretation of
experiments in terms of the Fermi surface and the variation of 
$\Delta({\bf k})$ on the Fermi surface. 
At this stage we only deal with conventional superconductors with the 
usual $s$-wave pairing. As it happens this is the only case for which reliable data already exists. 
Within the limits of a number of simplifying assumptions the above 
semiclassical theory provides for Bohr-Sommerfeld like
quantisation rules for quasiparticles. In particular it allows for the 
analogue of magnetic breakdown which involves tunnelling between
distinct semiclassical orbits. We show that there are orbits which enclose
areas that are precisely the same size as the Landau orbits in the normal
state, but involve such tunnelling.
As will be seen the consequence of these tunnelling events is a damping
factor in the Lifshitz-Kosevich formulae in agreement with experiments
\cite{onuki:92:0,janssen:98:0,goll:96:0}.

The theory we wish to use seeks the semiclassical spectrum of the following
Bogoliubov-de Gennes (BdG) equations
\begin{equation}
	\left(
		\begin{array}{cc}
			\frac{1}{2m}
			\left(
				\hat{\bf p} +e{\bf A}({\bf r})
			\right)^2
			+ V({\bf r}) -\epsilon_F
			&
			\!\!\!\!\!\!\!\!\!\!\!\!\!\! |\Delta({\bf r})| e^{i\phi({\bf r})} \\
			|\Delta({\bf r})| e^{-i\phi({\bf r})} &
			\!\!\!\!\!\!\!\!\!\!\!\!\!\! -\frac{1}{2m}
			\left(
				\hat{\bf p} -e{\bf A}({\bf r})
			\right)^2
			- V({\bf r}) +\epsilon_F
		\end{array}
	\right)
	\left(
		\begin{array}{c}
			u_\lambda ({\bf r}) \\
			v_\lambda ({\bf r}) 
		\end{array}
	\right)
	=
	E_\lambda
	\left(
		\begin{array}{c}
			u_\lambda ({\bf r}) \\
			v_\lambda ({\bf r}) 
		\end{array}
	\right),
\label{eqn:BdGequations}	
\end{equation}
where $u_\lambda({\bf r})$ and $v_\lambda({\bf r})$ are the probability 
amplitudes that an elementary excitation is a quasiparticle and quasihole
respectively, $\Delta({\bf r})=|\Delta|e^{i\phi({\bf r})}$ is the order 
parameter and the other symbols have the conventional meaning. For Type II
superconductors in large magnetic fields $\Delta({\bf r})$ takes the form of
the Abrikosov flux lattice~\cite{abrikosov:88:0}, comprising a periodic 
array of vortices.
The effective classical mechanics, Hamilton-Jacobi equations for the
quasiparticles, corresponding to (\ref{eqn:BdGequations}), is described
by the following effective Hamiltonians \cite{duncan:02:0}:
\begin{equation}
	E^\alpha({\bf p},{\bf r})
	=
	{\bf p}\cdot{\bf v}_s({\bf r})
	+\alpha
	\sqrt{
		\left(
			\frac{p^2}{2m}+\frac{1}{2}mv_s^2({\bf r}) 
				+ V({\bf r}) -\epsilon_F
		\right)^2
		+ |\Delta({\bf r})|^2
	},
\label{eqn:ClassicalHamiltonian}
\end{equation}
where $\alpha=\pm$ and 
$m{\bf v}_s({\bf r})=\frac{1}{2} \hbar {\mathbf \nabla} \phi({\bf r})+e{\bf A}({\bf r})$ is the superfluid velocity.
On a constant energy surface these determine, implicitly, the functions
${\bf p}^\alpha({\bf r})$ to be used in the Bohr-Sommerfeld quantisation
rule. To simplify matters we take the crystal potential $V({\bf r})$ to be
a constant.
It is clear from the single vortex solution~\cite{duncan:02:0}, that the
detailed shape of $|\Delta({\bf r})|$ is not essential. This fact will be
used below. By contrast the r\^ole of 
the line of phase singularities~\cite{duncan:02:0}, which runs along the vortex
core, needs more careful consideration. The
topologically non-trivial behaviour of the phase gradient 
($\oint {\mathbf \nabla}\phi \cdot d{\bf r}\neq 0$ for any path enclosing 
lines of phase singularities) causes the phase gradient, 
${\mathbf \nabla}\phi$, to cancel out (on the average) the increase in
${\bf A}({\bf r})$ across the cell. Consequently the
Abrikosov form for ${\bf v}_s({\bf r})$ is periodic. The Hamiltonian, 
(\ref{eqn:ClassicalHamiltonian}), then describes quasiparticles which 
correspond to spinor Bloch waves \cite{franz:00:0}.
On the other hand the topologically non-trivial behaviour of
$\Delta({\bf r})$ arises from the superposition of topologically {\it trivial}
solutions to the Ginzburg-Landau theory, so long as the coefficients are
suitably determined from minimization of the free energy for the nonlinear 
problem\cite{abrikosov:88:0}. We can therefore consider the solution of the
BdG equations for a topologically trivial 
$\Delta({\bf r})=|\Delta(y)|\exp(ik_xx)$, with 
$|\Delta(y)|=\Delta_0 
\exp(  
	-\frac{1}{2}(y-\frac{\hbar k_x}{2eB})^2/(2\pi)^{-1}d^2
)$ and 
$k_x=2\pi n/d$ ($d$ the flux lattice cell size, $n$ an integer). 
A full Bloch wave solution to equation (\ref{eqn:BdGequations}) 
can then be constructed by forming an appropriate superposition 
and determining the coefficients from self-consistency. 
The single-valuedness of each solution ensures the resulting superposition
is also single-valued, and the former requires
the quantisation of a Bohr-Sommerfeld integral for quasiparticles with the
$\Delta({\bf r})$ given above.

Taking this approach, and working in the Landau gauge,
we rewrite (\ref{eqn:ClassicalHamiltonian}) as
\begin{equation}
	E^\alpha({\bf p},{\bf r})
	=
	P_x v_{s,x}(y)
	+\alpha
	\sqrt{
		\left(
			\frac{p^2}{2m}
			+\frac{1}{2}m v_{s,x}(y)
		 	-\epsilon_F
		\right)^2
		+ |\Delta(y)|^2
	    },
\label{eqn:ClassicalHamiltonian2}
\end{equation}
where $mv_{s,x}(y)=\hbar k_x/2 +eA_x(y)$ and
we have approximated
${\bf B}({\bf r})=\nabla \times {\bf A}({\bf r})$ by a constant (averaged) 
${\bf B}$-field. The quasiparticle dynamics described by 
(\ref{eqn:ClassicalHamiltonian2}) are integrable and we can deploy the full
semiclassical machinery developed in \cite{duncan:02:0}. This we now do.

The momentum branches defined by the Hamilton-Jacobi equations
$E^\alpha({\bf p},{\bf r})=E^\alpha$ are 
\begin{equation}
	p_y^{\pm,\alpha}(y)
	=
	\sqrt{
		p_F^2-P_x^2-p_z^2
		-m^2\omega_c^2 \left(
					\textstyle{\frac{\hbar k_x}{2eB}}-y
				\right)^2
		\pm 2m
		\sqrt{
			\left(
				E^\alpha 
				+P_x \omega_c 
				\left( 
					y - \textstyle{\frac{\hbar k_x}{2eB}}
				\right)
			\right)^2
			-|\Delta(y)|^2
		     }
	     }.
\label{eqn:pyFullPx}
\end{equation}
We see immediately that $\frac{\hbar k_x}{2eB}$ behaves like an orbit 
centre, though restricted to $\frac{\hbar k_x}{2eB}=nd$ ($n$ integer). Less
obviously, $P_x$, also behaves like an orbit centre (see below). It is 
natural to restrict it to one flux cell, {\it i.e.,} 
$-\frac{d}{2}\leq \frac{P_x}{eB} \leq +\frac{d}{2}$. Then, since 
$\frac{P_x}{eB}$ lies within a cell, and $\frac{\hbar k_x}{2eB}$ represents
translations by lattice cell vectors, the pair combined allow us to centre
orbits anywhere in the sample. 
As we will see, different placements of orbits within a cell lead to different
$E_\lambda$, {\it i.e.,} degeneracy within a cell is lifted. 
On the other hand 
translations of such an orbit by lattice vectors (different choices of $k_x$)
leave $E_\lambda=E_\lambda(P_x)$ invariant.
One further simplification replaces $|\Delta(y)|$ by $\langle |\Delta(y)|
\rangle=|\Delta|$ in the region of interest. Outside this region we expect
to set $|\Delta(y)|=0$ but, since from (\ref{eqn:pyFullPx}), any choice 
in the range $0 \leq \Delta(y) \leq |\Delta|$ for $y-\hbar k_x/2eB > d$
makes little difference, we can for simplicity still take 
$|\Delta(y)|=|\Delta|$. Taking all these observations into account
the constant energy orbits corresponding to these relations are the same as
those studied by Burmistrov and Dubovskii \cite{burmistrov:96:0}. 
Firstly we investigate the orbits and spectrum for $P_x=0$.
\begin{figure}
	\onefigure[width=8cm,clip=]{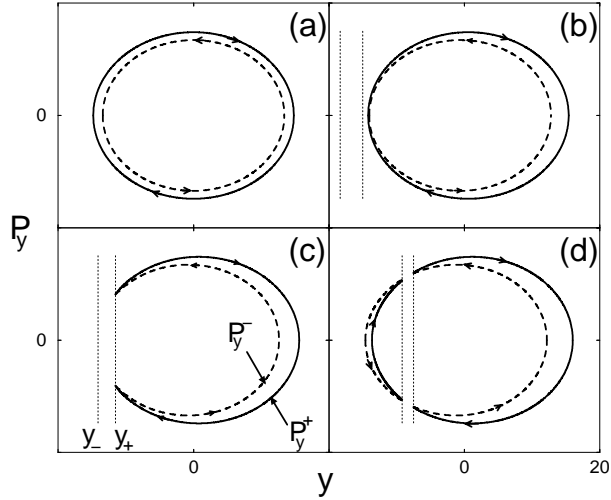}
	\caption{Phase space orbits defined by $p_y^{\pm,\alpha}$ given in 
		(\protect{\ref{eqn:pyFullPx}}) with $|\Delta(y)|=|\Delta|$. 
		The solid line is for a 
		particle-like excitation, the dashed line is for a hole-like 
		one. Arrows indicate the direction of the velocity. 
		{\bf (a)} $P_x=0$. The turning points along the $y$-axis are 
		given by $p^{\pm,\alpha}_y=0$. {\bf (b)} $P_x \neq 0$ and 
		small. The orbits are shifted in opposite directions. 
		The vertical dotted lines indicate the position of the 
		Andreev-like turning points, $y_+(P_x)$ and $y_-(P_x)$, see 
		equation (\protect{\ref{eqn:MASPoints}}). {\bf (c)} As $P_x$ 
		increases further the Andreev-like turning point $y_+$, for 
		which $p_y^{\pm,\alpha} \neq 0$, is reached before the normal 
		turning points given by $p_y^{\pm,\alpha}=0$. MAS takes place
		(see text).
		 {\bf (d)} $P_x$ increases even further. Now MAS also occurs at
		 $y_-$ creating a second particle-hole orbit. The two orbits 
		are separated in real space by 
		$\delta y =2|\Delta|/P_x \omega_c$.}
	\label{fig:Orbits}
\end{figure}
Fig.~\ref{fig:Orbits}(a)  shows typical classical phase space orbits defined
by (\ref{eqn:pyFullPx}) with $P_x=0$. The corresponding Bohr-Sommerfeld
quantisation condition yields
\begin{equation}
	E^\pm_n(p_z)
	=
	\pm
	\sqrt{
		\left(
			\hbar \omega_c \left(
						n+\frac{1}{2}
					\right)
			+\frac{p_z^2}{2m}
			-\epsilon_F
		\right)^2
		+
		|\Delta|^2
	     } \ .
\label{eqn:SpectrumPxZero}
\end{equation}
This is the Landau level spectrum, shifted into two square root singularities
(Fig.~\ref{fig:LandauLevelSpectrum}), which was studied previously by Miller
and Gy\"orffy~\cite{miller:95:0} and Miyake~\cite{miyake:93:0}. 
As it turns
out this theory is over simplified but nevertheless it contains the basic
physics. The picture is one of Landau levels which march across the 
gap~\cite{miller:95:0} giving the same frequency as in the normal state, 
but with
an extra damping, due to $|\Delta|$ \cite{miller:95:0,miyake:93:0}, given
by $R^{sc}=aK_1(a)$, with $a=2\pi \ell \frac{|\Delta|}{\hbar \omega_c}$,
($\ell$ integer), where $K_1$ is the Bessel function for imaginary argument.
Whilst $R^{sc}(a)$ fits the experimental data it predicts too much damping
for realistic values of $|\Delta|$. However, as was pointed 
out in ref.~\cite{burmistrov:96:0} $P_x \neq 0$ changes the spectra drastically
because, as is clear from (\ref{eqn:ClassicalHamiltonian}) `Landau level'
energies depend upon $P_x$. The rest of this paper deals with these 
complications.

Thus we consider the orbits and spectrum for $P_x \neq 0$. 
Figure~\ref{fig:Orbits}(b), (c) and (d) demonstrate the marked change in the
structure of the phase space orbits as $P_x$ increases from zero 
(figure~\ref{fig:Orbits}(a)). To understand the behaviour of the branches
of $p_y^{\pm,\alpha}$, it is helpful to view the orbit structure as resulting
from the competition between two different types of turning point. The usual
turning points, $p_y^{\pm,\alpha}=0$, are due to the harmonic oscillator
confining potential, $\frac{1}{2}m\omega_c^2 \tilde{y}^2$, 
($\tilde{y}=y-\hbar k_x/2eB$) in (\ref{eqn:pyFullPx}),
and are those seen in figure~\ref{fig:Orbits} (a) and (b). However the
presence of the second square root in (\ref{eqn:pyFullPx}),
$\sqrt{(E+P_x\omega_c \tilde{y})^2 - |\Delta|^2}$, which we denote by 
$\epsilon(\tilde{y})$ 
provides an alternative reflection mechanism for which 
$p_y^{\pm,\alpha} \neq 0$. If $\epsilon(\tilde{y})$ becomes zero before 
the normal
turning points are reached the momentum becomes complex and the wave function
will correspondingly exhibit evanescent decay. Classically the excitation 
undergoes reflection. These new turning points given by 
$\epsilon(\tilde{y})=0$ are
\begin{equation}
	y_\pm
	=
	\frac{\pm |\Delta| -E}{P_x \omega_c}.
\label{eqn:MASPoints}
\end{equation}
This reflection process is analogous to Andreev scattering~\cite{andreev:64:0}.
Andreev scattering is due to scattering from inhomogeneities in 
$|\Delta({\bf r})|$ Thus $\sqrt{E^2 -|\Delta({\bf r})|^2} \rightarrow 0$ as
$|\Delta({\bf r})| \rightarrow E$ from below. The reflection process we 
consider has $|\Delta|=constant$, but due to the non-zero vector potential
the extra term, $P_x \omega_c \tilde{y}$, 
can still cause $\epsilon(\tilde{y}) \rightarrow 0$. (It is for this reason
that the detailed shape of $|\Delta(y)|$ is not essential.)
To emphasise the similarity of this process with Andreev reflection, and the 
r\^ole of the vector potential we shall call the scattering mechanism
Magnetic Andreev Scattering (MAS), and $y_\pm$ Andreev-like turning points.

The appearance of MAS when $P_x$ is varied (Fig.~\ref{fig:Orbits}(c) and (d))
can then be explained as follows. From (\ref{eqn:MASPoints}) we see that as
$P_x \rightarrow 0$, $y_\pm \rightarrow -\infty$ ($E\geq |\Delta|$). When
$P_x$ becomes non-zero the Andreev-like turning points move in from $-\infty$
towards the normal turning points. We also see, as stated above,
that $P_x$ acts essentially
like an orbit centre, the particle-like and hole-like orbits (see below) 
being pulled in {\it opposite} directions due to their opposite charge 
(Fig.~\ref{fig:Orbits}(b)). The orbits first intersect when 
$p^+_y(y_+)=p^-_y(y_+)$ {\it i.e.,} when the Andreev-like 
turning point reaches the
normal turning point. For a more physical picture of the nature of these 
turning points remember that the spinor (the eigenfunction to 
(\ref{eqn:BdGequations})) being transported along a given trajectory has both
particle, $|u|$, and hole, $|v|$, amplitudes and thus carries an effective
charge $e^*=e(|u({\bf r})|^2-|v({\bf r})|^2)$. Evidently $e^*({\bf r})$ is
a function of position and in particular changes from $e^*>0$ along
$p_y^+$ (Fig.~\ref{fig:Orbits}(c)), through zero (at the turning point)
to $e^*<0$ along $p_y^-$. So the effective charge of the excitation changes
sign at $y_+$ and correspondingly the direction of circulation in the 
magnetic field is reversed. A particle-like excitation is reflected as a 
hole-like excitation, and a given orbit has both particle-like and hole-like
segments.

Having discussed the appearance and interpretation of the MAS orbits we now
turn to their quantisation. The spectrum obtained from semiclassical 
quantisation is shown in Fig.~\ref{fig:PxNonZeroSpectrum}. 
\begin{figure}
	\twofigures[width=7cm,clip=]{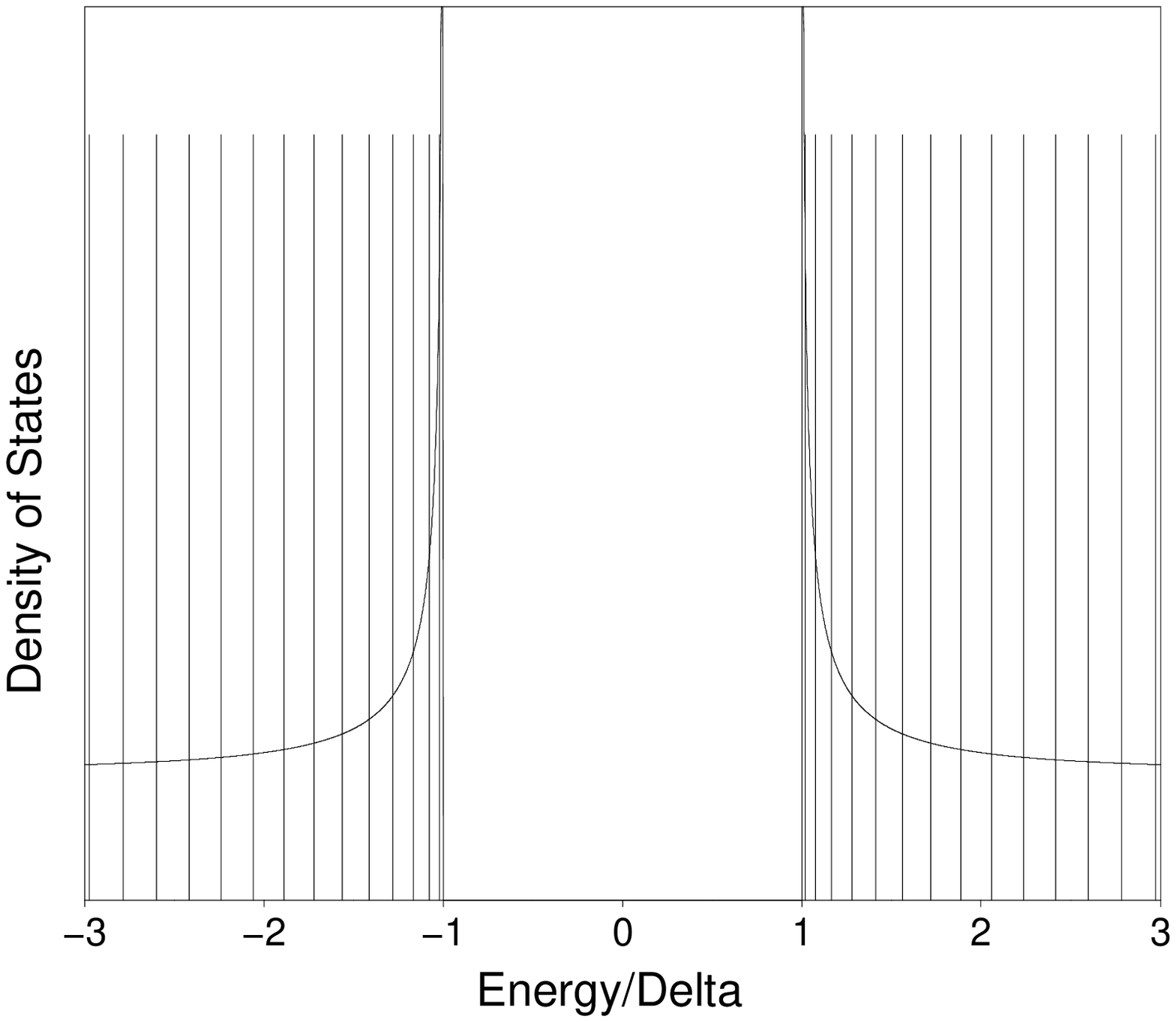}{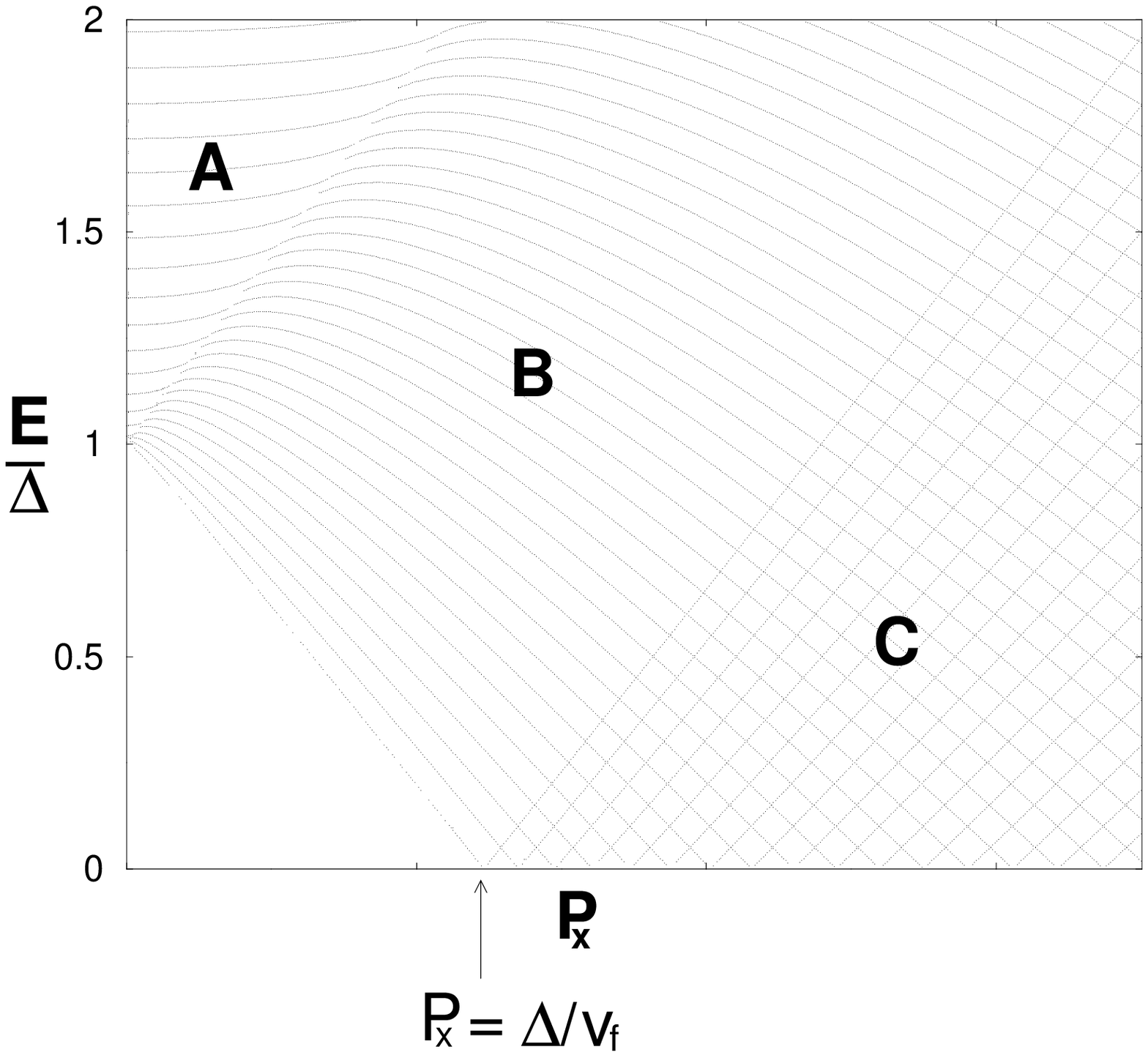}
	\caption{Landau level spectrum defined by \protect{(\ref{eqn:SpectrumPxZero})}. The BCS density of states is included to emphasise how it is broken upin a magnetic field into Landau levels, which are nonetheless pushed apart by the gap, $|\Delta|$.}
	\label{fig:LandauLevelSpectrum}
	\caption{Semiclassical spectrum obtained by quantising the orbits in Fig.~\protect{\ref{fig:Orbits}}. Region A corresponds to quantising the orbits in Fig.~\protect{\ref{fig:Orbits}}(a) and (b), whilst region B and C correspond to quantising the one or two scattering orbits in Fig.~\protect{\ref{fig:Orbits}}(c) and (d). The spectrum first becomes gapless when $P_x=p_{min}\approx |\Delta(B)|/v_F$.}
	\label{fig:PxNonZeroSpectrum}
\end{figure}
Firstly note that
at $P_x = 0$ there are Landau levels corresponding to 
(\ref{eqn:SpectrumPxZero}) which are excluded from the gap. As we 
increase $P_x$ we reach the situation shown in Fig.~\ref{fig:Orbits}(c). The
sensitive dependence of the turning point $y_+(P_x)$ upon $P_x$ results in a
spectrum which is highly $P_x$ dependent. Increasing $P_x$ further results
in the situation shown in Fig.~\ref{fig:Orbits}(d). The dramatic new feature
of the spectrum for $P_x \neq 0$ is the existence of states inside the gap.
These states are new features of the theory which were not there in the 
$P_x=0$ case. However despite the existence of these states it is clear that
they cannot on their own account for the experimental facts. These states 
have considerably different phase space (and hence momentum space) area, so
that the dHvA frequency for each individual orbit, which is related to this
area, will be very different from that in the normal state. To explain 
the experiments we in fact require one further bit of physics. That is
a phenomenon analogous to `Magnetic Breakdown', \cite{duncan:99:0,duncan:99:1} 
which we shall now discuss.

\begin{figure}[h]
	\twofigures[width=6.8cm,clip=]{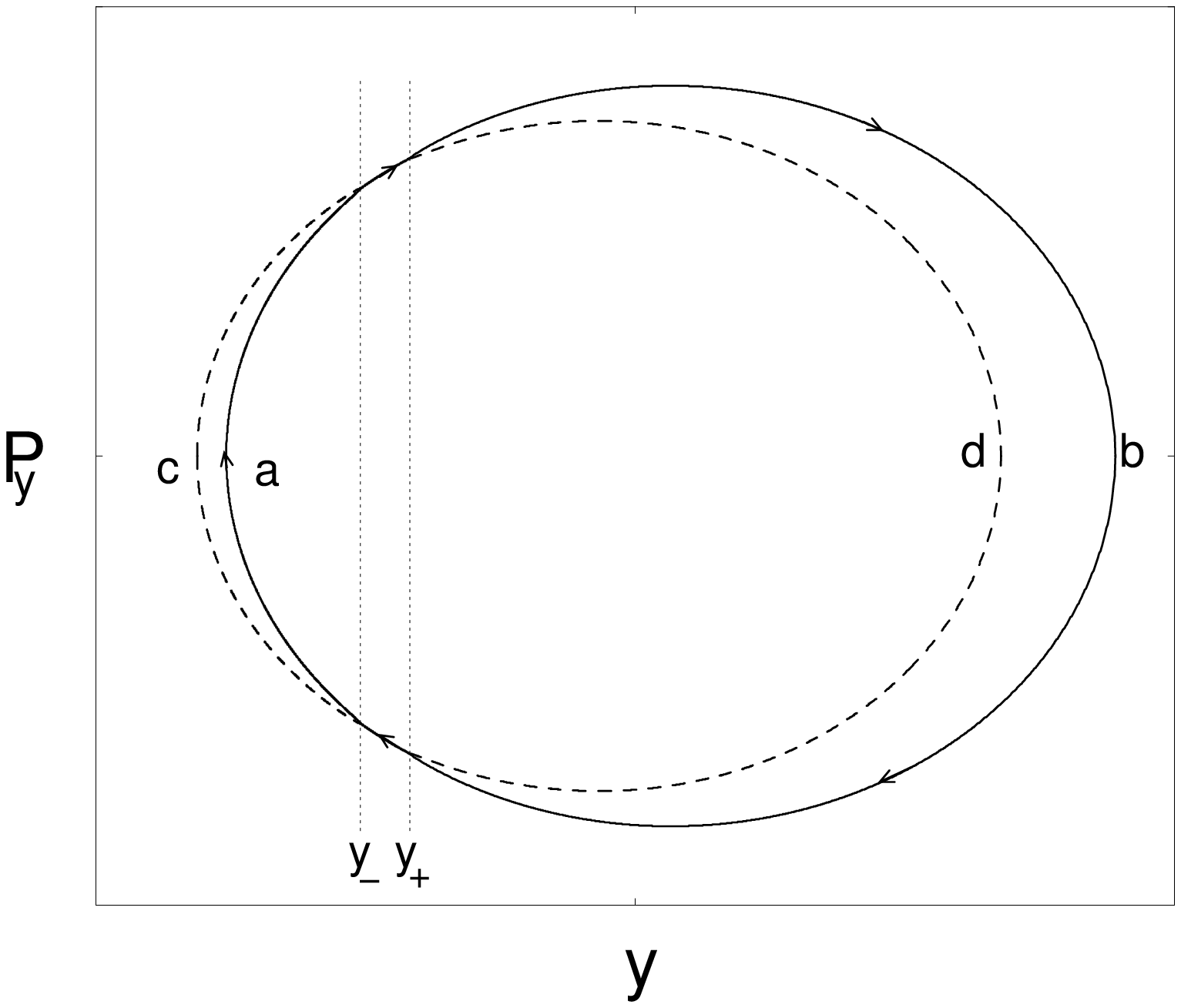}{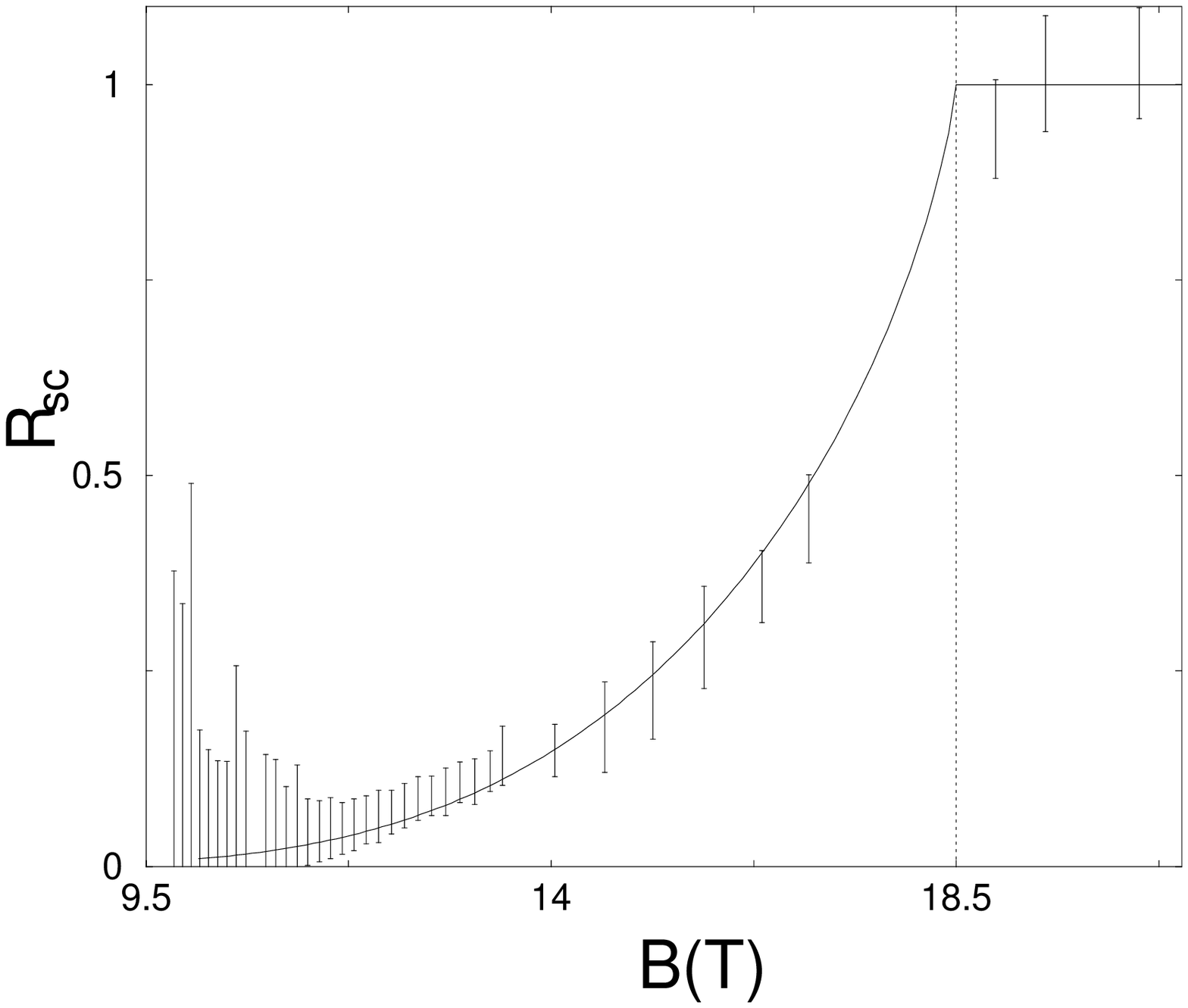}
	\caption{The tunnelling orbit (solid line) comprised of segments from both of the MAS orbits. Tunnelling from $y_+$ to $y_-$ results in an orbit with precisely the same phase space area as for an orbit in the normal state. The spectrum quantising such an orbit is given in equation (\protect{\ref{eqn:TunnellingSpectrum}}).}
	\label{fig:TunnellingOrbit}
	\caption{Fit of the current theory for the extra damping, $R^{sc}$, in the superconducting state to the data for V$_3$Si using $|\Delta|=4.4$ meV. The upper critical field is B$_{c2}$=18.5T.}
	\label{fig:TheoryAndData}
\end{figure}

We observed that the separation of the two orbits in Fig.~\ref{fig:Orbits}(d)
is given by $\delta y = y_+ -y_- = 2|\Delta|/P_x \omega_c$ and it thus
inversely proportional to both $P_x$ and the ${\bf B}$-field (through
$\omega_c =eB/m$). For a fixed $B$, as $P_x$ increases this separation in 
real space rapidly shrinks and quasiparticle tunnelling becomes a viable 
option. Purely on physical grounds we see that tunnelling between orbit
segments (Fig.~\ref{fig:TunnellingOrbit}) can reproduce an orbit in phase
space with approximately the normal area. 
To quantise explicitly the tunnelling
orbit the Bohr-Sommerfeld integral can be represented as a complex contour
integral. The Riemann surface for the problem is somewhat similar to that of
Fortin {\it et al}~\cite{fortin:98:0}. In our case the MAS orbits live on two
different sheets of the Riemann surface but the tunnelling orbit remains on
only one sheet. Consequently the integral is equal to the residue at $\infty$, 
on this sheet, and we find 
\begin{equation}
	E_n(p_z)
	=
	\hbar \omega_c \left(
				n+\frac{1}{2}
			\right)
	+\frac{p_z^2}{2m}
	-\epsilon_F.
\label{eqn:TunnellingSpectrum}
\end{equation}
The spectrum is $P_x$-independent and the tunnelling orbits have exactly the
normal state frequency. However the tunnelling coefficient, which depends upon
the imaginary part (Im) of the momentum, is 
\begin{equation}
	T(P_x)
	=
	e^{
	    -\frac{1}{\hbar} \left|
			        \int_{y_-(P_x)}^{y_+(P_x)}dy' 
				\mathrm{Im} \ p_y^\pm(y')
			     \right|
	  },
\label{eqn:TunnellingFactor}
\end{equation}
and clearly depends upon $P_x$ and $B$. For fixed $P_x$, 
$\delta y \propto 1/B \rightarrow 0$ as $B$ increases, so that the possibility
of tunnelling is enhanced. We can therefore view the tunnelling as
`Magnetic Breakdown of MAS orbits'. Magnetic Breakdown in the normal state is
a well studied phenomena~\cite{fortin:98:0}. The oscillatory magnetisation
formulae is modified due to the tunnelling. To account for this we must 
include the factor $T^{2k}$ (every revolution of the orbit 
we pick up 2 tunnelling 
coefficients, and in general we have $k$ revolutions). The $P_x$ dependence
of (\ref{eqn:TunnellingFactor}) requires the $\sum_{P_x}T^{2k}$ to be 
carried out when calculating the magnetisation. (An additional $\sum_{k_x}$
yields the usual degeneracy factor in the formulae for the magnetisation.)
Observing that $P^2_{x,\mathrm{max}}\ll p_F^2$ we find 
$\hbar^{-1} \int dy' \mathrm{Im} \ p_y^\pm(y') \approx \frac{\pi}{2}
\frac{|\Delta|^2}{\hbar \omega_c}\frac{1}{v_F}\frac{1}{P_x}$, where $v_F$ is
the Fermi velocity. 
Using $|\Delta(B)|=\Delta(0)(1-B/B_{c2})^{1/2}\rightarrow 0$ as 
$B\rightarrow B_{c2}$ we have the important observation that 
$T(P_x)\rightarrow 1$ for all $P_x$ as $B \rightarrow B_{c2}$. Thus the
additional damping vanishes as $|\Delta|\rightarrow 0$ as is to be expected.
Note, tunnelling is {\it not} a small correction, it is zeroth order. The
normal state is recovered for $T(P_x) \rightarrow 1$.

The main contribution to $\sum_{P_x}T^{2k}(P_x)$ is
obtained for the largest value of $P_x=eBd/2$ and is given by
\begin{equation}
	T^{2k}(P_x=eBd/2)
	=
	e^{
	    -2^{\frac{3}{2}}\pi^{\frac{1}{2}} k
		\frac{|\Delta|^2}{\hbar \omega_c}\frac{\Lambda}{v_F}
	  },
\label{eqn:TunnellingMax}
\end{equation}
where $\Lambda=(2eB\hbar)^{-1/2}$. Interestingly this result is very 
similar to that of Wasserman and Springford~\cite{wasserman:96:0}. More
generally, the extra damping factor in the superconducting state is found
to be 
\begin{equation}
	R^{sc}(B)
	=
	\frac{1}{p_d}\int_{p_{min}(B)}^{p_d(B)}
	dP_x
	e^{
	    -\frac{2k}{\hbar} \left|
			      	\int_{y_-(P_x,B)}^{y_+(P_x,B)}
				dy'\mathrm{Im}p_y^\pm(y',P_x)
			      \right|
	  },
\label{eqn:DampingFactor}
\end{equation}
where $p_d=eBd/2$, and $p_{min}(B)\cong |\Delta(B)|/v_F$ is the smallest
value of $P_x$ for which the spectrum is gapless 
(see Fig.~\ref{fig:PxNonZeroSpectrum}). Although the above results have 
been derived for a free electron model with short range attraction, 
constant (averaged) pairing potential ($|\Delta|$) and uniform
magnetic field ${\bf B}$, provisionally the formulae for the damping factor
given in (\ref{eqn:DampingFactor}) can be compared with that deduced from
experiments. We replace $v_F$ by its orbitally averaged quantity, and use
$\langle 1/v_F \rangle = m_b/\sqrt{2\hbar e F}$ with $m_b=0.9m_e$, 
$F=1560$T, and $|\Delta|=4.4$ meV to compare (\ref{eqn:DampingFactor})
with the data for V$_3$Si in figure \ref{fig:TheoryAndData}. 
Even at this
early stage we have a satisfactory fit to the data. Clearly, $|\Delta|$ in 
the above analysis is an effective quantity whose value cannot be expected
to agree with the zero field $|\Delta|$ of 2.6 meV (see for example
ref.~\cite{janssen:98:0}). A more detailed analysis of the existing data as 
well as a more complete presentation of the theory will be given 
elsewhere~\footnote{work in preparation}.

To summarise we developed a semiclassical theory in the superconducting state.
We consider new semiclassical orbits involving `magnetic breakdown' of MAS
orbits. These new orbits, when quantised, have precisely the normal state
frequency. The tunnelling involved gives rise to an extra damping in the
superconducting state in agreement with experiments in such classic 
superconductors as V$_3$Si. When generalised to include period crystal
potential, a more complete description of the flux lattice, and anisotropic
pairing~\cite{mineev:99:0}, the above theory will also be suitable for
analysing experimental results on exotic superconductors when ever these
become available.

\acknowledgments
We would like to thank Steve Hayden for supplying us with the data for
V$_3$Si in Fig.~\ref{fig:TheoryAndData}. This work was supported by
EPSRC, grant No. GR/M53844.

\end{document}